\documentclass[12pt,preprint]{aastex}

\shorttitle{The dynamics of internal working surfaces in MHD jets}
\shortauthors{De Colle et al.}

\begin{document}

\title{The dynamics of internal working surfaces in MHD jets}

\author{Fabio De Colle}
\affil{Dublin Institute for Advanced Studies (DIAS), 31 Fitzwilliam Place, Dublin 2, Ireland}
\email{fdc@cp.dias.ie}

\and

\author{Alejandro C. Raga and Alejandro Esquivel}
\affil{Instituto de Ciencias Nucleares, UNAM, A. Postal 70-543,
04510 M\'exico, M\'exico}
\email{raga, esquivel@nucleares.unam.mx}

\begin{abstract}

The dynamical effects of magnetic fields in models of radiative,
Herbig-Haro (HH) jets have been studied in a number of papers. For example,
magnetized, radiative jets from variable sources have been studied
with axisymmetric and 3D numerical simulations.
In this paper, we present an
analytic model describing the effect of a toroidal magnetic field
on the internal working surfaces that result from a variability in
the ejection velocity. We find that for parameters appropriate for
HH jets the forces associated with the magnetic field dominate over
the gas pressure force within the working surfaces. Depending on
the ram pressure radial cross section of the jet, the magnetic
field can produce a strong axial pinch, or, alternatively, a
broadening of the internal working surfaces. We check the validity
of the analytic model with axisymmetric numerical simulations of
variable, magnetized jets.

\end{abstract}



\section{Introduction}

It is now relatively certain that some Herbig-Haro (HH) jets have knot structures
which are the result of a time-variability in the ejection. For
example, the observations of some jets with organized structures
of knots of different sizes (e.~g., HH 30, 34 and 111, see \citealt{esq07}, 
\citealt{rag02} and \citealt{mas02}) can be reproduced surprisingly well
with variable ejection jet models. In the present paper, we study
the effect of the presence of a magnetic field on the evolution of a variable jet.

It is still an open question to what extent magnetic fields are
important in determining the dynamics of HH jets. The associated
problem of radiative, MHD jets has been explored in some
detail in the existing literature.
\citet{cer97}, and \citet{cer99} computed 3D simulations of radiative,
MHD jets with different magnetic field configurations (at the injection
point). \citet{fra98} carried out axisymmetric simulations of
similar flows.

The problem of an MHD, radiative jet ejected with a time-variable velocity
was explored with axisymmetric simulations by \citet{gara00,
garb00, sto00, osu00, fra00, dec06} and \citet{har07}.
Variable, MHD jets were also explored with 3D simulations by
\citet{cer01a,cer01b}. The general conclusions
that can be obtained from these simulations is that the internal working
surfaces produced by the ejection variability are not affected strongly
by a poloidal magnetic field. On the other hand, if the magnetic field
is toroidal (or, alternatively, has a strong toroidal component), the
material within the working surfaces of the jet flow has a stronger
concentration towards the jet axis.

\citet{gara00} showed that in a variable ejection velocity
jet the ``continuous jet beam'' sections in between the working surfaces
have a low toroidal magnetic field, which grows in strength quite
dramatically when the material goes through one of the working surface
shocks into one of the knots. In the present paper, we present a
simple, analytic model from which we obtain the conditions under
which the toroidal magnetic field produces an axial compression
of the internal working surfaces. This analytic model is presented
in \S 2. In \S 3, we present axisymmetric numerical simulations in
which we compare the working surfaces with and without a toroidal magnetic
field, showing the effect described by the analytic model. Finally,
in \S 4 we present our conclusions.

\section{The radial motion of the material within an internal working surface}

\subsection{General considerations}

A time-variability in the ejection velocity leads to the formation
of two-shock ``internal working surfaces'' which travel down the
jet flow. In a frame of reference that moves with the working surface,
the flow takes the configuration shown in Figure 1, with material
entering the shocked layer from both the upstream and downstream
directions.

Let us consider an internal working surface within a cylindrically
symmetric jet with a toroidal magnetic field configuration.
The material in the jet beam cross section or within the
working surface is subject to two radial forces: the magnetic pinch force
\begin{equation}
F_m=-{B\over 4\pi r}{d\over dr} (rB)\,,
\label{fm}
\end{equation}
where $B$ is the toroidal magnetic field and $r$ the cylindrical radius,
and the force due to the pressure gradient
\begin{equation}
F_p=-{dP\over dr}\,,
\label{fp}
\end{equation}
where $P$ is the gas pressure.
The cross section of the jet is in lateral equilibrium when
$F=F_m+F_p=0$, it will be subject to a lateral expansion
when $F>0$ and to a compression when $F<0$.

Let us now assume that the jet beam has a generic cross section of
the form
\begin{equation}
\rho(r)=\rho_0 {\underline \rho}(r)\,,
\label{rho}
\end{equation}
\begin{equation}
B(r)=B_0 {\underline B}(r)\,,
\label{b}
\end{equation}
\begin{equation}
v(r)=v_0 {\underline v}(r)\,,
\label{v}
\end{equation}
where $\rho(r)$ is the density, $B(r)$ the magnitude of the
(toroidal) magnetic field, and $v(r)=v_j-v_{w}$ is the relative velocity
with which the jet material (moving at a velocity $v_j$)
enters the working surface (which moves at a velocity $v_{w}$),
see Figure 1.
The constants $\rho_0$, $B_0$ and $v_0$ correspond to characteristic
values of the respective quantities, and ${\underline \rho}(r)$,
${\underline B}(r)$ and
${\underline v}(r)$ are dimensionless functions of the radius giving the
radial dependence of the flow variables from $r=0$ (the symmetry
axis) out to $r=r_j$ (the outer radius of the jet beam). In principle,
these three dimensionless functions are of order one
unless very strong changes in the flow variables occur across
the jet cross section.

Let us now consider that the material goes through the ``Mach disk''
shock of an internal working surface. If we assume that the shock
is strong (i.~e., that it is highly supersonic and superalfv\'enic),
from the Rankine-Hugoniot equations for MHD (e.~g. \citealt{dra93})
the post-shock radial cross section is given by~:
\begin{displaymath}
P_w^{nr}(r)={2\over {\gamma+1}}\rho_0 v_0^2 {\underline \rho}(r)
{\underline v}^2(r)\,;
\end{displaymath}
\begin{equation}
P_w^{rad}(r)={(8\pi)^{1/2} \rho_0^{3/2} v_0 c_w^2\over B_0}
{{\underline \rho}^{3/2}(r){\underline v}(r)\over {\underline B}(r)}\,,
\label{pw}
\end{equation}
\begin{displaymath}
\rho_w^{nr}(r)={{\gamma+1}\over{\gamma-1}}\rho_0
{\underline \rho}(r)\,;
\end{displaymath}
\begin{equation}
\rho_w^{rad}(r)={(8\pi)^{1/2} \rho_0^{3/2} v_0\over B_0}
{{\underline \rho}^{3/2}(r){\underline v}(r)\over {\underline B}(r)}\,,
\label{rhow}
\end{equation}
\begin{displaymath}
B_w^{nr}(r)={{\gamma+1}\over{\gamma-1}}B_0
{\underline B}(r)\,;
\end{displaymath}
\begin{equation}
B_w^{rad}(r)={(8\pi)^{1/2} \rho_0^{1/2} v_0}
{{\underline \rho}^{1/2}(r){\underline v}(r)}\,,
\label{bw}
\end{equation}
where $P_w^{nr}(r)$, $\rho_w^{nr}(r)$ and $B_w^{nr}(r)$ are
the post-shock gas pressure, density and magnetic field cross sections
for the case of a non-radiative shock, and 
$P_w^{rad}(r)$, $\rho_w^{rad}(r)$ and $B_w^{rad}(r)$ are
the cross sections for the case of a radiative shock
in which the post-shock gas instantaneously cools to an
isothermal sound speed $c_w$. As we have said above, equations
(\ref{pw}-\ref{bw}) have been derived for the case of a strong
shock. In order to obtain these relations it is also necessary
to assume that the pre-shock Alfv\'enic Mach number has values
smaller than $\sim M_w^2=(v/c_w)^2$.

The factors including the
specific heat ratio $\gamma$ (see equations \ref{pw}-\ref{bw}) take
the numerical values $2/(\gamma+1)=3/4$ and $(\gamma+1)/(\gamma-1)=4$
for the case of a monoatomic, non-relativistic gas (i.~e., for
$\gamma=5/3$). From now on, we will use these numerical values in
order to simplify the equations.

Combining equations (\ref{pw}-\ref{bw}) with (\ref{fm}-\ref{fp}) we
obtain the magnetic and gas pressure forces acting radially on the
post-Mach disk material. The resulting magnetic force is
\begin{equation}
F^{nr}_m={4 B_0^2\over \pi r_j} f^{nr}_m(r)\,; \qquad
F^{rad}_m={2\rho_0 v_0^2\over r_j}
f^{rad}_m(r)\,,
\label{Fm}
\end{equation}
where the dimensionless force $f_m(r)$ is given by
\begin{displaymath}
f^{nr}_m(r)=-{\underline B}(r){r_j\over r}{d\over dr}
\left[r{\underline B}(r)\right]\,;
\end{displaymath}
\begin{equation}
f^{rad}_m(r)=-{\underline \rho}^{1/2}(r){\underline v}(r){r_j\over r}
{d\over dr}
\left[r{\underline \rho}^{1/2}(r){\underline v}(r)\right]\,.
\label{ffm}
\end{equation}
The resulting gas pressure force is
\begin{equation}
F^{nr}_p={3\rho_0 v_0^2\over 4 r_j}f^{nr}_p(r)\,;\qquad
F^{rad}_p={(8\pi)^{1/2} \rho_0^{3/2} {v_0} c_w^2\over r_j B_0}
f^{rad}_p(r)\,,
\label{Fp}
\end{equation}
where the dimensionless force $f_p(r)$ is given by
\begin{equation}
f^{nr}_p(r)=-r_j{d\over dr}\left[{\underline \rho}(r)
{\underline v}^2(r)\right] \,;
\qquad
f^{rad}_p(r)=-r_j{d\over dr}\left[{{\underline \rho}^{3/2}(r)
{\underline v}(r)\over {\underline B}(r)}\right]\,.
\label{ffp}
\end{equation}

\subsection{Scaling properties of the magnetic and gas pressure forces}

Let us now consider the ratio $M/P$ between the moduli of the
magnetic and gas pressure forces. From equations (\ref{Fm}) and (\ref{Fp})
we obtain
\begin{equation}
(M/P)_{nr}={64\over 3 M_A^2}\left|{f_m^{nr}(r)\over f_p^{nr}(r)}\right| \,;
\qquad
(M/P)_{rad}={\sqrt{2} M_w^2\over {M_A}}
\left|{f_m^{rad}(r)\over f_p^{rad}(r)}\right|\,,
\label{mp}
\end{equation}
where $M_A\equiv v_0/v_A$ is the Alfv\'enic Mach number (obtained with
the characteristic velocity $v_0$ and the Alfv\'en velocity
$v_A=B_0/\sqrt{4\pi \rho_0}$), $M_w=v_0/c_w$ is the sonic Mach number (calculated
with the characteristic velocity $v_0$ and the post-shock sound
speed $c_w$ of the radiative shock) and
the $f_m(r)$ and $f_p(r)$ functions are given by equations
(\ref{ffm}) and (\ref{ffp}), respectively.

One can argue that if the dimensionless cross section of the
jet (described by equations \ref{rho}-\ref{v}) is smooth, then
the $f_m^{nr}(r)$, $f_m^{rad}(r)$ and $f_p(r)$ functions
(see equations \ref{ffm} and \ref{ffp}) will have values of
order 1.

In our derivation of the pressure force within the internal
working surface, we have only considered the gradient of the
post-shock gas pressure. Of course, the fact that the working
surface material is free to leave through the sides of the jet
beam will lead to an extra gas pressure gradient (directed outwards),
particularly in the case of a non-radiative flow. The dimensionless
pressure cross section due to this effect is still likely to lead
to a dimensionless force $f_p(r)\sim 1$.

Setting $f_m^{nr}(r),
f_m^{rad}(r),\,f^{nr}_p(r),\,f^{rad}_p(r)\sim 1$, from equation
(\ref{mp}) we then obtain
\begin{equation}
(M/P)_{nr}\sim {64\over 3 M_A^2} \,;
\qquad
(M/P)_{rad}\sim {M_w^2\over {M_A}}\,.
\label{mpp}
\end{equation}
From these two estimates of the ratio between the magnetic and gas
pressure forces, we conclude that
\begin{itemize}
\item for the non-radiative case~: if the Alv\'enic Mach number of
the flow entering the Mach disk is large (e.~g., $M_A>10$) we have
$(M/P)_{nr}\ll 1$, and therefore the lateral expansion or contraction
of the gas within the working surface will be governed by the gas
pressure force,
\item for the radiative case~: if we consider jets with given values for
$v_A$ and $c_w$, it is clear that as the velocity $v_0$ increases,
the $(M/P)_{rad}$ ratio increases (proportional to ${v_0}$). In
particular, if we have flows with $v_A\sim c_w$, the magnetic
to gas pressure force ratio has values $(M/P)_{rad}\sim M_w$. Thus,
for a Mach disk in the strong shock regime, the post-shock magnetic
pressure force will under most conditions dominate over the gas pressure
force.
\end{itemize}
Therefore, for the non-radiative and the radiative cases, whether
the jet material within the working surface expands or contracts in
the radial direction will be determined by the signs of $f_p^{nr}(r)$
and $f_m^{rad}(r)$, respectively (see equations \ref{ffm} and \ref{ffp}),
provided that the Mach number of the jet has values $M_w\sim 10$ or
larger.

\section{Simulations of the internal working surface of an HH jet}

Let us now consider the case of a jet model with a
``top hat'' density and velocity initial cross section, and an initial
toroidal magnetic field cross section of the form
\begin{equation} 
B(r)=B_0{r\over r_j}\,.
\label{ba}
\end{equation}
This kind of magnetic field cross section has been used in many of
the previous simulations of radiative MHD jets (see, e.~g., \citealt{gara00}).
 With this cross section for the jet beam, we have
\begin{equation}
f^{nr}_m=-2r/r_j\,,\qquad f^{rad}_m=-r_j/r\,,
\label{fm1}
\end{equation}
and
\begin{equation}
f^{nr}_p=0\,, \qquad f^{rad}_p=\left({r_j\over r}\right)^2\,.
\label{fp1}
\end{equation}
In other words, the magnetic pressure force is directed towards
the axis, and the gas pressure force (acting in the radial direction
on the working surface jet material) is zero for the non-radiative
case, and points outwards for the radiative case.

We now compute models of a jet with this initial cross section,
and an initial scale of the magnetic field $B_0=0$ (i.~e., a purely
hydrodynamic jet) and $B_0=5\,\mu$G.
The jet is injected with a constant density $n_j=100$~cm$^{-3}$,
temperature $T_j=900$~K and radius $r_j=2\times 10^{15}$~cm,
and moves into a homogeneous, unmagnetized
environment of density $n_{env}=10$~cm$^{-3}$ and temperature
$T_{env}=9000$~K. The injection velocity varies sinusoidally
with time, with a period $\tau=20$~yr, a half-amplitude of
$150$~km~s$^{-1}$, and an average velocity of 300~km~s$^{-1}$.

For the two chosen values of the magnetic field ($B_0=0$ and 5~$\mu$G,
see above and Equation \ref{ba}), we run both non-radiative simulations
and simulations in which we include the coronal ionization equilibrium
cooling function of \citet{dal72}. 
These simulations are
run with the uniform grid, axisymmetric MHD code described in detail by
\citet{dec06}.
The codes uses a second order up-wind scheme, which integrates the MHD 
equations using a Godunov method with a Riemann solver. 
The Riemann problem is solved using primitive variables and the 
magnetic field divergence is maintained close to zero using the 
CT method \citep{tot00}.
The computational
domain of $(5,1)\times 10^{16}$~cm (axial, radial) extent is resolved
with $2000\times 400$ grid points. A reflection condition is applied
on the jet axis and on the $z=0$ plane in the $r>r_j$ region. An outflow
condition is applied in the remaining grid boundaries.

The time-dependent ejection velocity of the jet leads to the formation
of successive internal working surfaces that travel down the jet flow.
It is possible to estimate the ratio $(M/P)$ between the magnetic and pressure
forces within the internal working surfaces by noting that the shock velocity
(associated with the two working surface shocks) has a value
$v\approx 150$~km~s$^{-1}$. In other words, the value of the shock
velocity is of the order of the half-amplitude of the ejection velocity
variability (see, e.~g., \citealt{rag90}).

With this value of $v$ and the initial jet density and
temperature, we can compute $M_A=v/v_A\approx 31$ (where
$v_A=4.8$~km~s$^{-1}$ for our $B_0=5\,\mu$G value
and our initial jet density),
$M_w=150$ (for an assumed post-cooling sound speed of 10~km~s$^{-1}$)
and then we use equation (\ref{mp}) to obtain
$(M/P)_{nr}\sim 0.02$ and $(M/P)_{rad}\sim 10^3$. Therefore,
the magnetic force should have little effect in the non-radiative
simulations, and result in similar structures for the internal working surfaces
in the cases of magnetized and non-magnetized jets.

Figure 2 shows that our numerical simulations do show this
effect. In this Figure, we show the
density stratification obtained for non-radiative jets with $B_0=
0$ (left) and $B_0=5\,\mu$G (right) after a $t=90$~yr time-integration.
It is clear that though the details of the flow are affected by
the presence of a toroidal magnetic field, the general features
of the two working surfaces produced within the computational domain
are quite similar in the magnetized and non-magnetized cases.

The fact that $(M/P)_{rad}\sim 700$ (see above) implies that the
magnetic pinch force should dominate the dynamics of the material
within the internal working surfaces. Our two radiative numerical
simulations (shown in Figure 3) do show this effect. In the
magnetized simulation, the internal working surfaces become
more strongly compressed towards the jet axis as they evolve
(traveling away from the source), an effect not seen in the
non-magnetized, radiative jet simulation (see Figure 3).

Figure 4 shows zooms of the knot situated at
$z\approx 3.5\times 10^{16}$~cm (the knot on the top
half of the $t=90$~yr time frames shown in Figures 1 and 2) for
our four computed models. This Figure shows that in the
non-radiative case, the two working surface shocks
have a separation which is similar to the diameter
of the jet, and that the density structures are
very similar for the $B_0=$ and 5~$\mu$G models.

As expected, much higher densities are
obtained in the radiative jet simulations. In the
radiative case, the working surface obtained from the
$B_0=5$~$\mu$G model shows larger densities, a
much higher concentration towards the jet axis and
larger separations between the working surface
shocks than the $B_0=0$ model. 

More complex profiles for the magnetic field and the pressure 
were explored in the past by several authors 
(e.g. \citealt{gara00, garb00, sto00, osu00, fra00, dec06}),
with results similar to the one obtained by the simple configurations
of magnetic field and pressure presented here.
As shown in Section 2.2, the expansion or contraction of
the material in the working surface is nearly independent on the 
initial profile of the pressure. On the other side, the pre-shock magnetic field
profile contributes to $f^{rad}_p$ (but not to $f^{rad}_m$), and to the value of
$(M/P)$. For small values of the magnetic field (e.g. close to the jet axis)
$(M/P) \lesssim 1$, and the gas pressure force dominates.

Also, we note that \citet{beg98} has studied the development of pinch instabilities 
in non radiative jets, as due to the presence of a toroidal magnetic field. 
He found that a condition necessary for the development of the pinch instability is 
\begin{equation}
 \frac{d \ln B}{d \ln r} > \frac{\gamma \beta -2}{\gamma \beta +2},
\end{equation}
where $\beta = 8 \pi P/B^2$.
In the case of a radiative working surface with a post-shock region 
with $\beta \ll 1$ (corresponding to the condition $(M/P)_{rad} \gg 1$)
this condition reduces to $F_m < 0$.

\section{Conclusions}

It is a known result that internal working surfaces in radiative,
MHD jets with a toroidal magnetic field configuration form dense,
axial structures, which do not appear in unmagnetized jets. We
present a simple, analytic model with which we show that the strong
jump conditions (applied to one of the working surface shocks)
imply that the magnetic force dominates over the gas pressure force
within a radiative working surface and that the gas pressure force
is dominant for a non-radiative working surface (provided that one
has a shock Mach number of at least $M_w\sim 10$ and an Alv\'enic Mach
number which does not exceed $M_w^2$).

Interestingly, the radial dependence
of the toroidal magnetic field within a radiative working surface
depends only on the cross section of the pre-shock ram pressure
$p_{ram}(r)=\rho(r)v^2(r)$ impinging on the shocks. From equation (\ref{ffm}),
we can see that if we have a $p_{ram}(r)$ that decreases towards
the edge of the jet faster than $1/r^2$, the magnetic force within the
working surface will be directed outwards, and will tend to increase
the width of the working surface.

We have run four simulations (with a top hat cross section for $p_{ram}$,
that results in an axially directed magnetic pinch within the working surfaces),
therefore in complete consistency with our analytic model. We find
that in the non-radiative case the presence of a toroidal magnetic
field has very little effect on the structure of the internal working
surfaces. We also find that for the radiative case, the presence of
a toroidal magnetic field produces a strong axial compression of the
material within the internal working surfaces (see Figure 4).

The analytic model presented in this paper can then be used
to decide what ram pressure and toroidal magnetic field 
cross section to use in a magnetized, radiative, variable jet simulation 
in order to produce internal working surfaces that show narrower or 
broader structures than what is obtained in non-magnetized jet simulations. 
This might be a valuable tool when trying to model the knots in specific HH jets,
and might provide a possible method for constraining the strength and
the configuration of magnetic fields within such objects.

\acknowledgments
AR and AE acknowledge support from
the DGAPA (UNAM) grant IN108207, from the CONACyT grants 46828-F and
61547, and from the ``Macroproyecto de Tecnolog\'\i as para la Universidad
de la Informaci\'on y la Computaci\'on (Secretar\'\i a de Desarrollo
Institucional de la UNAM). This work is supported in part by the European
Community's Marie Curie Actions - Human Resource and Mobility within
the JETSET (Jet Simulations, Experiments and Theory) network under
contract MRTN-CT-2004 005592.
We thank Enrique Palacios, Mart\'\i n Cruz and Antonio Ram\'\i rez
for supporting the servers in which the calculations of this paper
were carried out.

\clearpage

\begin{figure}
  \epsscale{.60}
  \plotone{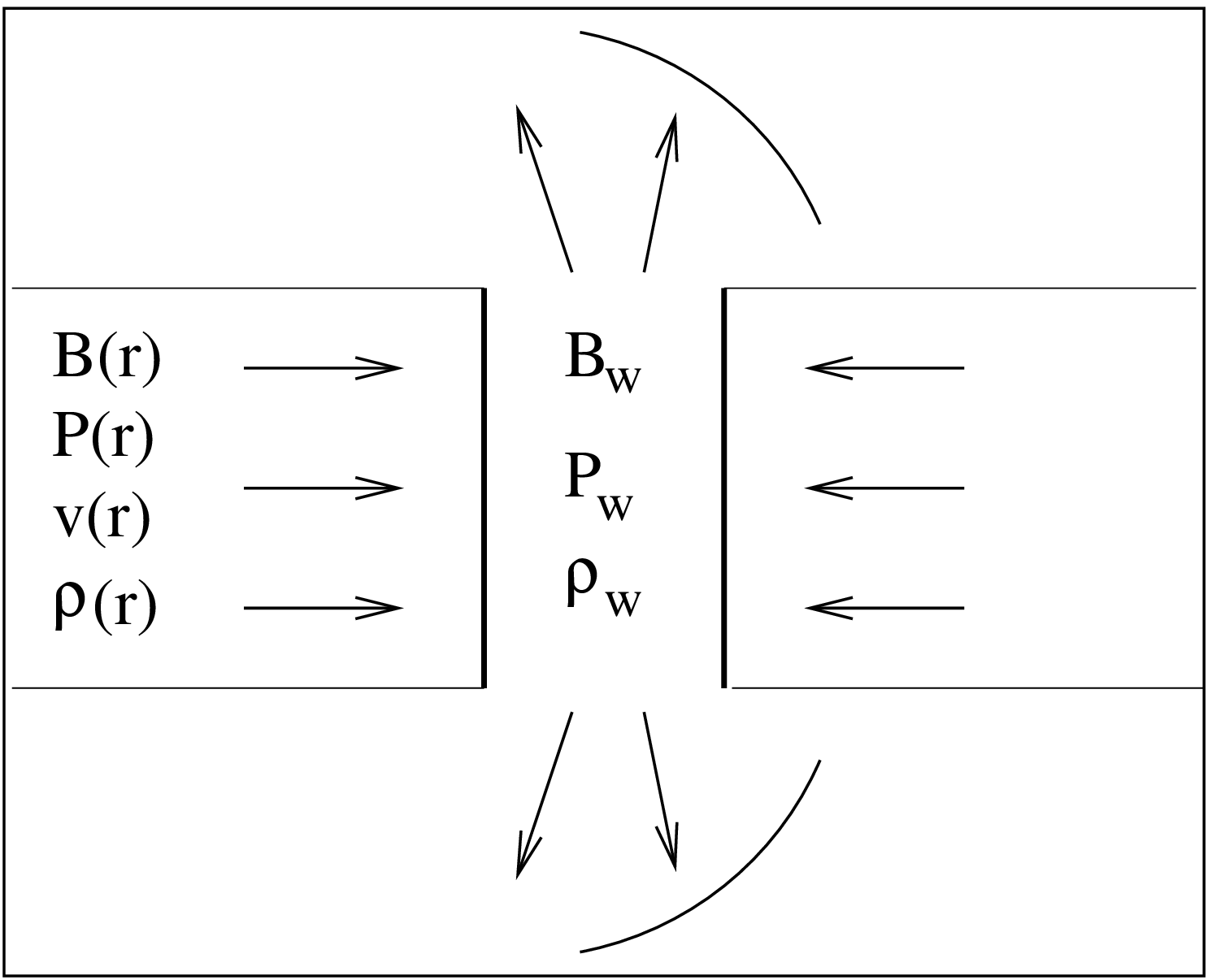}
  \caption{Schematic diagram showing an internal working surface
produced by an ejection velocity variability (the source is to the
left). In a frame of
reference that moves with the velocity $v_{ws}$ of the working surface,
material from the continuous jet segments enters the shocked layer
from both the upstream and downstream directions. The $B(r)$,
$P(r)$, $v(r)$, $\rho(r)$ radial cross section of the pre-shock
region produces a $B_w(r)$, $P_w(r)$, $\rho_w(r)$ cross section
within the shocked layer (in this shocked layer, the velocity along
the jet axis
is $\sim 0$ in the reference frame moving with the working surface).
The material in the working surface exits laterally, shocking against
the jet cocoon.}
  \label{f1}
\end{figure}

\clearpage

\begin{figure}
  \epsscale{.55}
  \plotone{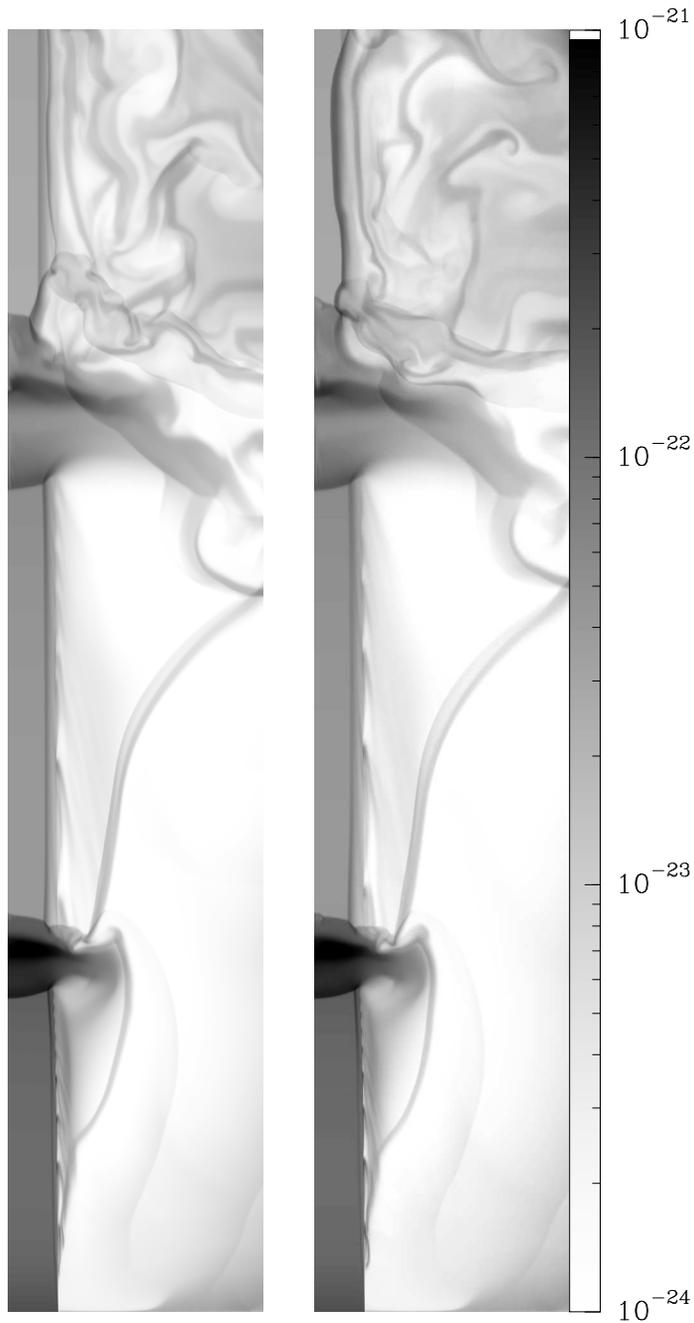}
  \caption{Density structures resulting from $t=90$~yr time
integrations of the non-magnetized (left) and magnetized (right)
non-radiative jet models. The flow is injected from the bottom
of the grid, and travels upwards, producing internal working
surfaces (two of these are seen in the displayed time frames).
The frames cover the full, $(5,1)\times 10^{16}$~cm (axial, radial)
computational domain. The densities are shown with the logarithmic
gray scale given (in g~cm$^{-3}$) by the bar on the right.}
  \label{f2}
\end{figure}

\clearpage

\begin{figure}
  \epsscale{.55}
  \plotone{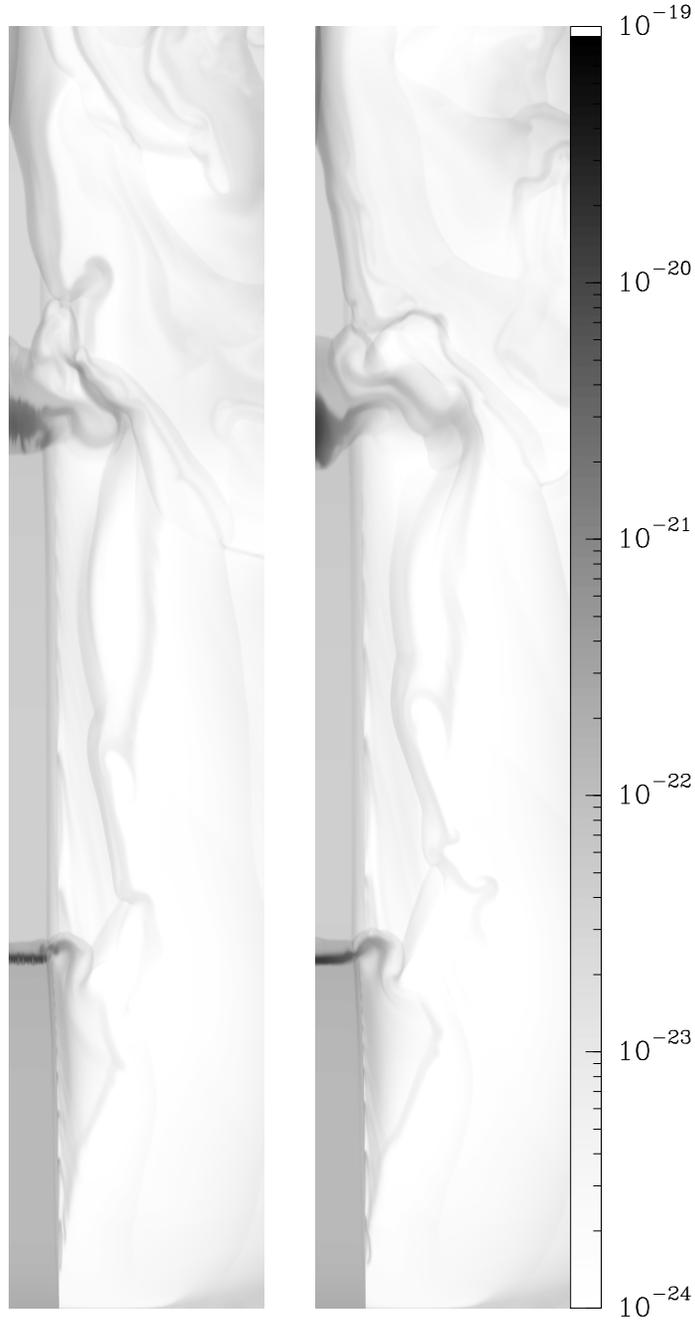}
  \caption{Density structures resulting from $t=90$~yr time
integrations of the non-magnetized (left) and magnetized (right)
radiative jet models.
The frames cover the full, $(5,1)\times 10^{16}$~cm (axial, radial)
computational domain. The densities are shown with the logarithmic
gray scale given (in g~cm$^{-3}$) by the bar on the right.}
  \label{f3}
\end{figure}

\clearpage

\begin{figure}
  \epsscale{.60}
  \plotone{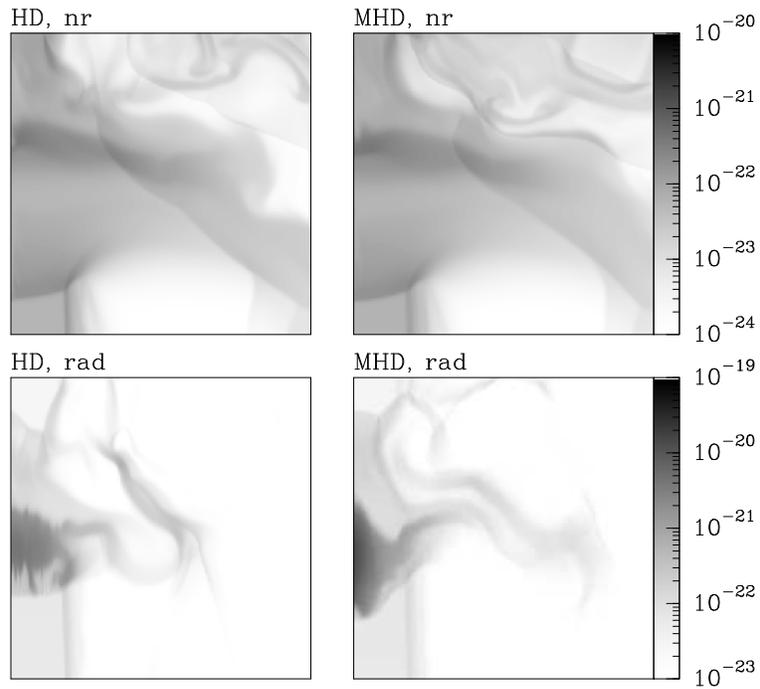}
  \caption{Density stratifications of the knot seen in the
upper half of the $t=90$~yr time frames shown in Figures 2 and 3.
The non-radiative models are shown on the top, and the radiative
models on the bottom. The non-magnetized models are on the left,
and the magnetized ones on the right. The displayed domain
has an axial and radial size of $7.5\times 10^{15}$~cm. The
density of the non-radiative models is given (in g~cm$^{-3}$)
by the bar on the top right, and the density of the radiative
models by the bar on the bottom right.}
  \label{f4}
\end{figure}

\end{document}